\documentclass[11pt,showpacs,showkeys,preprintnumbers,amsmath,amssymb]{revtex4}

\begin{document}
% You may use Title,Subject,Author,Manager,Company,Operator,
% Category,Comment,Hlinkbase document properties here
\title{Unheeded pseudo solution of Dirac-Coulomb equations with an indirect
transformation of functions}
\author{Ruida Chen³Âï}

\affiliation{Shenzhen Institute of Mathematics and Physics,
Shenzhen, 518028, China}

%\date{\today}

\begin{abstract}
We open out one of incorrect solutions of the Driac equation in the
Coulomb field given in a published paper. By introducing a
transformation of function, the paper transformed the original
radial first-order Dirac-Coulomb equation into two second-order
Dirac-Coulomb equation. However, each of the second-order
differential equations has differential energy eigenvalues set. The
original paper wrote the two differential equations into one of
forms, and then gave the distinguished energy eigenvalues. The
mathematical procedure is not correct. For the same quantum system,
introducing a transformation of function yields two different energy
eigenvaluse, the result violates the uniqueness of solution. It
actually shows that the given second-order differential equations
have no solution. On the other hand, the given formal solutions of
the second-order Dirac-Coulomb equations violate the conditions for
determining solution. Consequently, the solutions given by the
author are pseudo solution, and the corresponding energy eigenvalues
set is also a pseudo eigenvalues set.

\end{abstract}



\maketitle

\section{Introduction}

It is well known that, in mathematics, for a differential equation of
high-order or system of differential equations of first-order, only the
formal solutions satisfy the original equations, satisfy the conditions for
determining solution and satisfy the uniqueness of solution or the
eigensolutions set can be regard as the correct solutions. However, in
relativistic quantum mechanics, it seems that many literatures are ignorant
of such fundamental mathematics rules. For example, some published papers
derived two different second-order Dirac-Coulomb equations for the two
components of wave functions which has actually different eigenvalues sets,
but finally wrote the two second-order differential equations in one form of
second-order differential equations by introducing some pretexts such as
so-called ``decoupling
procedure''\cite{Martin:1958}\cite{Biedenharn:1962}, mentioning lightly,
enshrouded the contradictions of violating uniqueness of solutions. In
particular, those given formal solutions are not in agreement with the
boundary condition. They are divergent at the origin of the coordinate
system. Whereas, only one use the boundary condition can determine the
solution of the Draic-Coulomb equations of second order. On the other hand,
in order to obtain a new energy eigenvalues, some papers even introduced the
so-called position-dependent
mass\cite{Alhaidari:2004}\cite{Vakarchuk:2005} to spelled backward
second-order Dirac-Coulomb\cite{Chen:1} equation, and all of the
corresponding results are not the necessary mathematical deductions,
including the second-order differential equations.

In the present paper, we disclose one of hidden pseudo solutions of the
Dirac equations in Coulomb field, which was published in 2005. The paper
first introduced a transformation of functions for the two components of
wave functions, and transformed the original first-order Dirac equation in
the Coulomb field into two second-order Dirac-Coulomb equations. The two
second-order differential equations haave different formal eigenvalues set,
violating the uniqueness of solutions, and the formal eigenfunctions are
also divergent at the origin of coordinate system. But the paper wrote the
two differential equations into one form, having concealed the mathematical
contradiction. The paper also fenced with the boundary condition for the
complete eigenfunctions. We conclude that the formula of energy-levels which
corresponds to the eigenvlue set is a pseudo energy eigenvalues set for the
hydrogen and hydrogen-like atom, and the given formal solutions are the
pseudo solutions of the Dirac-Coulobm equations.

\section{One of second-order Dirac-Coulomb equation and its pseudo solutions}

Discussing the single-electron or single-muon atoms for which the potential
produced by the nucleus can be well approximated by a central potential
$V\left( r \right)$, a paper\cite{Deck:2005} derived the coupled Dirac
equations with assuming the central potential has the Coulomb form $V\left(
r \right) = - {Ze^2} \mathord{\left/ {\vphantom {{Ze^2} {4\pi \varepsilon _0
r}}} \right. \kern-\nulldelimiterspace} {4\pi \varepsilon _0 r}$ exterior to
the nucleus as follows
\begin{equation}
\label{eq1}
\begin{array}{l}
 \left( {\frac{d}{dr} + \frac{1 + k}{r}} \right)g\left( r \right) -
\frac{1}{\hbar c}\left[ {E + mc^2 - V\left( r \right)} \right]f\left( r
\right) = 0 \\
 \left( {\frac{d}{dr} + \frac{1 - k}{r}} \right)f\left( r \right) +
\frac{1}{\hbar c}\left[ {E - mc^2 - V\left( r \right)} \right]g\left( r
\right) = 0 \\
 \end{array}
\end{equation}

\noindent
By introducing a mathematical transformation of functions
\begin{equation}
\label{eq2}
\rho = 2\sqrt {\frac{m^2c^4 - E^2}{\hbar ^2c^2}} r,\quad u = \rho
^{\frac{3}{2}}\left( {g - \sqrt {\frac{mc^2 + E}{mc^2 - E}} f} \right),\quad
v = \rho ^{\frac{3}{2}}\left( {g + \sqrt {\frac{mc^2 + E}{mc^2 - E}} f}
\right)
\end{equation}

\noindent
the system of first-order differential equations (\ref{eq1}) with a Coulomb
potential was transformed into two second-order Dirac equation for the
components $f$ and $g$
\begin{equation}
\label{eq3}
\begin{array}{l}
 \left\{ {\frac{d^2}{d\rho ^2} - \frac{1}{\rho ^2}\left[ {\left( {j +
\frac{1}{2}} \right)^2 - \left( {\alpha Z} \right)^2 - \frac{1}{4}} \right]
+ \frac{1}{\rho }\left( {\frac{\alpha ZE}{\sqrt {m^2c^4 - E^2} } +
\frac{1}{2}} \right) - \frac{1}{4}} \right\}u = 0 \\
 \left\{ {\frac{d^2}{d\rho ^2} - \frac{1}{\rho ^2}\left[ {\left( {j +
\frac{1}{2}} \right)^2 - \left( {\alpha Z} \right)^2 - \frac{1}{4}} \right]
+ \frac{1}{\rho }\left( {\frac{\alpha ZE}{\sqrt {m^2c^4 - E^2} } -
\frac{1}{2}} \right) - \frac{1}{4}} \right\}v = 0 \\
 \end{array}
\end{equation}

\noindent
where $j = 1 \mathord{\left/ {\vphantom {1 2}} \right.
\kern-\nulldelimiterspace} 2,3 \mathord{\left/ {\vphantom {3 2}} \right.
\kern-\nulldelimiterspace} 2, \cdots $, $\alpha = {e^2} \mathord{\left/
{\vphantom {{e^2} {2\varepsilon _0 hc}}} \right. \kern-\nulldelimiterspace}
{2\varepsilon _0 hc} = {e^2} \mathord{\left/ {\vphantom {{e^2} {4\pi
\varepsilon _0 \hbar c}}} \right. \kern-\nulldelimiterspace} {4\pi
\varepsilon _0 \hbar c}$. According to the opinion of the papers, the
equations (\ref{eq3}) have the general form of Whittaker's
equation\cite{Whittaker:1}
\begin{equation}
\label{eq4}
\left[ {\frac{d^2}{d\rho ^2} - \frac{1}{\rho ^2}\left( {\gamma ^2 -
\frac{1}{4}} \right) + \frac{\beta }{\rho } - \frac{1}{4}} \right]M\left(
\rho \right) = 0
\end{equation}

\noindent
where
\begin{equation}
\label{eq5}
\gamma ^2 = \left( {j + \frac{1}{2}} \right)^2 - \left( {\alpha Z}
\right)^2,\quad \beta = \beta _\pm = \frac{\alpha ZE}{\sqrt {m^2c^4 - E^2}
}\pm \frac{1}{2}
\end{equation}

\noindent
And the corresponding eigenvalues set was given as follows
\begin{equation}
\label{eq6}
E = mc^2\frac{{n}' + \sqrt {\left( {j + 1 \mathord{\left/ {\vphantom {1 2}}
\right. \kern-\nulldelimiterspace} 2} \right)^2 - \left( {\alpha Z}
\right)^2} }{\sqrt {\left( {\alpha Z} \right)^2 + \left[ {{n}' + \sqrt
{\left( {j + 1 \mathord{\left/ {\vphantom {1 2}} \right.
\kern-\nulldelimiterspace} 2} \right)^2 - \left( {\alpha Z} \right)^2} }
\right]^2} },\quad {n}' = 0,1,2, \cdots
\end{equation}

\section{Eigenvalues set of equations (\ref{eq3}) violates uniqueness of solution}

We first notice that (\ref{eq3}) involve two different equations, but they were
written in one form by introducing some new sign (\ref{eq5}). According to the
optimum theorem\cite{Chen:2000} for differential equations or the
asymptotic solution and power series solution, the uncoupled equations (\ref{eq3})
have different eigenvalues set\cite{Chen:2003}. One can demonstrate that
the formula (\ref{eq6}) for the energy levels in the Coulomb field is not a unique
mathematical deduction of the system of differential equations (\ref{eq3}).
Because the original paper didn't give the detail steps of the deduction,
the differential eigenvalues set for the same quantum system, meaning the
obvious mathematical difficult and contradiction, are not usually taken
attention to. Why don't we direct solve the second-order equations (\ref{eq3})
without the sign (\ref{eq5}) to find the formulas of the energy levels? The
conditions for determining solution of the Dirac equations in Coulomb field
are usually written in the rough form without considering the size of the
atomic nucleus as same as in Schr\"{o}dinger theory
\begin{equation}
\label{eq7}
R\left( {r \to \infty } \right) = 0,\quad - \infty < R\left( {0 \le r <
\infty } \right) < \infty
\end{equation}

Noticing the second-order differential equations (\ref{eq3}) have the asymptotic
solutions, as $\rho \to \infty $, $u \approx e^{ - \frac{\rho }{2}}$ and $v
\approx e^{ - \frac{\rho }{2}}$ are in agreement with the boundary
condition. Now we find the following form of the solution
\begin{equation}
\label{eq8}
u = e^{ - \frac{\rho }{2}}\chi ,\quad v = e^{ - \frac{\rho }{2}}\sigma
\end{equation}

\noindent
We have
\begin{equation}
\label{eq9}
\begin{array}{l}
 \frac{du}{d\rho } = e^{ - \frac{\rho }{2}}\left( {\frac{d\chi }{d\rho } -
\frac{1}{2}\chi } \right),\quad \frac{d^2u}{d\rho ^2} = e^{ - \frac{\rho
}{2}}\left( {\frac{d^2\chi }{d\rho ^2} - \frac{d\chi }{d\rho } +
\frac{1}{4}\chi } \right) \\
 \frac{dv}{d\rho } = e^{ - \frac{\rho }{2}}\left( {\frac{d\sigma }{d\rho } -
\frac{1}{2}\sigma } \right),\quad \frac{d^2v}{d\rho ^2} = e^{ - \frac{\rho
}{2}}\left( {\frac{d^2\sigma }{d\rho ^2} - \frac{d\sigma }{d\rho } +
\frac{1}{4}\sigma } \right) \\
 \end{array}
\end{equation}

\noindent
Substituting (\ref{eq8}) and (\ref{eq9}) into the
equations (\ref{eq3}) yields
\begin{equation}
\label{eq10}
\begin{array}{l}
 \frac{d^2\chi }{d\rho ^2} - \frac{d\chi }{d\rho } - \left[ {\left( {j +
\frac{1}{2}} \right)^2 - \left( {\alpha Z} \right)^2 - \frac{1}{4}}
\right]\frac{1}{\rho ^2}\chi + \left( {\frac{\alpha ZE}{\sqrt {m^2c^4 - E^2}
} + \frac{1}{2}} \right)\frac{1}{\rho }\chi = 0 \\
 \frac{d^2\sigma }{d\rho ^2} - \frac{d\sigma }{d\rho } - \left[ {\left( {j +
\frac{1}{2}} \right)^2 - \left( {\alpha Z} \right)^2 - \frac{1}{4}}
\right]\frac{1}{\rho ^2}\sigma + \left( {\frac{\alpha ZE}{\sqrt {m^2c^4 -
E^2} } - \frac{1}{2}} \right)\frac{1}{\rho }\sigma = 0 \\
 \end{array}
\end{equation}

\noindent
For $\chi $ and $\sigma $ we make the power series ansatz
\begin{equation}
\label{eq11}
\chi = \sum\limits_{n = 0}^\infty {b_n } \rho ^{s_\chi + n},\quad \sigma =
\sum\limits_{n = 0}^\infty {d_n } \rho ^{s_\sigma + n}
\end{equation}

\noindent
and insert them into the differential equations (\ref{eq10}), giving
\begin{equation}
\label{eq12}
\begin{array}{l}
 \sum\limits_{n = 0}^\infty {\left( {s_\chi + n} \right)\left( {s_\chi + n -
1} \right)b_n } \rho ^{s_\chi + n - 2} - \sum\limits_{n = 0}^\infty {\left(
{s_\chi + n} \right)b_n } \rho ^{s_\chi + n - 1} \\
 - \left[ {\left( {j + \frac{1}{2}} \right)^2 - \left( {\alpha Z} \right)^2
- \frac{1}{4}} \right]\frac{1}{\rho ^2}\sum\limits_{n = 0}^\infty {b_n }
\rho ^{s_\chi + n} + \left( {\frac{\alpha ZE}{\sqrt {m^2c^4 - E^2} } +
\frac{1}{2}} \right)\frac{1}{\rho }\sum\limits_{n = 0}^\infty {b_n } \rho
^{s_\chi + n} = 0 \\
 \sum\limits_{n = 0}^\infty {\left( {s_\sigma + n} \right)\left( {s_\sigma +
n - 1} \right)d_n } \rho ^{s_\sigma + n - 2} - \sum\limits_{n = 0}^\infty
{\left( {s_\sigma + n} \right)d_n } \rho ^{s_\sigma + n - 1} \\
 - \left[ {\left( {j + \frac{1}{2}} \right)^2 - \left( {\alpha Z} \right)^2
- \frac{1}{4}} \right]\frac{1}{\rho ^2}\sum\limits_{n = 0}^\infty {d_n }
\rho ^{s_\sigma + n} + \left( {\frac{\alpha ZE}{\sqrt {m^2c^4 - E^2} } -
\frac{1}{2}} \right)\frac{1}{\rho }\sum\limits_{n = 0}^\infty {d_n } \rho
^{s_\sigma + n} = 0 \\
 \end{array}
\end{equation}

\noindent
A comparison of coefficients yields
\begin{equation}
\label{eq13}
\begin{array}{l}
 \left[ {\left( {s_\chi + n} \right)\left( {s_\chi + n - 1} \right) - \left(
{j + \frac{1}{2}} \right)^2 + \left( {\alpha Z} \right)^2 + \frac{1}{4}}
\right]b_n \\
 - \left[ {\left( {s_\chi + n - 1} \right) - \left( {\frac{\alpha ZE}{\sqrt
{m^2c^4 - E^2} } + \frac{1}{2}} \right)} \right]b_{n - 1} = 0 \\
 \left[ {\left( {s_\sigma + n} \right)\left( {s_\sigma + n - 1} \right) -
\left( {j + \frac{1}{2}} \right)^2 + \left( {\alpha Z} \right)^2 +
\frac{1}{4}} \right]d_n \\
 - \left[ {\left( {s_\sigma + n - 1} \right) - \left( {\frac{\alpha
ZE}{\sqrt {m^2c^4 - E^2} } - \frac{1}{2}} \right)} \right]d_{n - 1} = 0 \\
 \end{array}
\end{equation}

The above recurrence relations determine the coefficients of the power
series (\ref{eq11}). All appearance, the above relations have the initial value:
$b_{ - 1} = d_{ - 1} = 0$, and $b_0 \ne 0$, $d_0 \ne 0$. Put $n = 0$ (\ref{eq13}),
we obtain
\begin{equation}
\label{eq14}
\begin{array}{l}
 \left[ {s_\chi \left( {s_\chi - 1} \right) - \left( {j + \frac{1}{2}}
\right)^2 + \left( {\alpha Z} \right)^2 + \frac{1}{4}} \right]b_0 = 0 \\
 \left[ {s_\sigma \left( {s_\sigma - 1} \right) - \left( {j + \frac{1}{2}}
\right)^2 + \left( {\alpha Z} \right)^2 + \frac{1}{4}} \right]d_0 = 0 \\
 \end{array}
\end{equation}

\noindent
It gives $s_\chi = s_\sigma = 1 \mathord{\left/ {\vphantom
{1 2}} \right. \kern-\nulldelimiterspace} 2\pm \sqrt {\left( {j + 1
\mathord{\left/ {\vphantom {1 2}} \right. \kern-\nulldelimiterspace}
2} \right)^2 - \left( {\alpha Z} \right)^2} $. Combining
(\ref{eq2}), (\ref{eq8}), (\ref{eq11}) and making use of the roots
of (\ref{eq14}) gives the complete form of the formal solutions of
the original first-order Dirac equation in the Coulomb field
\begin{equation}
\label{eq15}
\begin{array}{l}
 f = \frac{1}{2}\sqrt {\frac{mc^2 - E}{mc^2 + E}} \rho ^{ - 1\pm \sqrt
{\left( {j + 1 \mathord{\left/ {\vphantom {1 2}} \right.
\kern-\nulldelimiterspace} 2} \right)^2 - \left( {\alpha Z} \right)^2} }e^{
- \frac{\rho }{2}}\left( {\sum\limits_{n = 0}^\infty {d_n } \rho ^n -
\sum\limits_{n = 0}^\infty {b_n } \rho ^n} \right) \\
 g = \frac{1}{2}\rho ^{ - 1\pm \sqrt {\left( {j + 1 \mathord{\left/
{\vphantom {1 2}} \right. \kern-\nulldelimiterspace} 2} \right)^2 - \left(
{\alpha Z} \right)^2} }\left( {\sum\limits_{n = 0}^\infty {d_n } \rho ^n +
\sum\limits_{n = 0}^\infty {b_n } \rho ^n} \right) \\
 \end{array}
\end{equation}

\noindent
Here we have replaced the original indirect transformation
of functions (\ref{eq2}) by the corresponding direct transformation
of functions
\begin{equation}
\label{eq16}
g = \frac{1}{2}\rho ^{ - \frac{3}{2}}\left( {v + u} \right),\quad f =
\frac{1}{2}\sqrt {\frac{mc^2 - E}{mc^2 + E}} \rho ^{ - \frac{3}{2}}\left( {v
- u} \right)
\end{equation}

\noindent
Since the wave functions have to satisfy the boundary
condition (\ref{eq7}) at the origin of the coordinate system we must
choose the positive sign for square root of index $s_\chi $ and
$s_\sigma $. For the negative square root solution it follows that
$f \sim \rho ^{ - 1 - \sqrt {\left( {j + 1 \mathord{\left/
{\vphantom {1 2}} \right. \kern-\nulldelimiterspace} 2} \right)^2 -
\left( {\alpha Z} \right)^2} }$, $g \sim \rho ^{ - 1 - \sqrt {\left(
{j + 1 \mathord{\left/ {\vphantom {1 2}} \right.
\kern-\nulldelimiterspace} 2} \right)^2 - \left( {\alpha Z}
\right)^2} }$ near $r = 0$ namely $\rho = 0$, which would yield a
divergent integral for the norm\cite{Greiner:2000}. Hence,
\begin{equation}
\label{eq17}
s_\chi = s_\sigma = \frac{1}{2} + \sqrt {\left( {j + 1 \mathord{\left/
{\vphantom {1 2}} \right. \kern-\nulldelimiterspace} 2} \right)^2 - \left(
{\alpha Z} \right)^2}
\end{equation}

\noindent
and the formal solutions for the original Dirac-Coulomb equations(\ref{eq1})
should be written in the form
\begin{equation}
\label{eq18}
\begin{array}{l}
 f = \frac{1}{2}\sqrt {\frac{mc^2 - E}{mc^2 + E}} \rho ^{ - 1 + \sqrt
{\left( {j + 1 \mathord{\left/ {\vphantom {1 2}} \right.
\kern-\nulldelimiterspace} 2} \right)^2 - \left( {\alpha Z} \right)^2} }e^{
- \frac{\rho }{2}}\left( {\sum\limits_{n = 0}^\infty {d_n } \rho ^n -
\sum\limits_{n = 0}^\infty {b_n } \rho ^n} \right) \\
 g = \frac{1}{2}\rho ^{ - 1 + \sqrt {\left( {j + 1 \mathord{\left/
{\vphantom {1 2}} \right. \kern-\nulldelimiterspace} 2} \right)^2 - \left(
{\alpha Z} \right)^2} }\left( {\sum\limits_{n = 0}^\infty {d_n } \rho ^n +
\sum\limits_{n = 0}^\infty {b_n } \rho ^n} \right) \\
 \end{array}
\end{equation}

\noindent
On the other hand, the wave functions also have to satisfy
the boundary as $r \to \infty $ namely $\rho \to \infty $ we must
cut off the two power series in (\ref{eq18}) in order to $g\left(
{\rho \to \infty } \right) = 0$, $f\left( {\rho \to \infty } \right)
= 0$. It is assumed that $b_{n_r } \ne 0$, $b_{n_r } \ne 0$ and
$b_{n_r + 1} = b_{n_r + 2} = \cdots = 0$, $d_{n_r + 1} = d_{n_r + 2}
= \cdots = 0$. Inserting $n = n_{r + 1} + 1$ in the recurrence
relations (\ref{eq13}) educes that
\begin{equation}
\label{eq19}
\begin{array}{l}
 \left( {\frac{1}{2} + \sqrt {\left( {j + 1 \mathord{\left/ {\vphantom {1
2}} \right. \kern-\nulldelimiterspace} 2} \right)^2 - \left( {\alpha Z}
\right)^2} + n_r } \right) - \left( {\frac{\alpha ZE}{\sqrt {m^2c^4 - E^2} }
+ \frac{1}{2}} \right) = 0 \\
 \left( {\frac{1}{2} + \sqrt {\left( {j + 1 \mathord{\left/ {\vphantom {1
2}} \right. \kern-\nulldelimiterspace} 2} \right)^2 - \left( {\alpha Z}
\right)^2} + n_r } \right) - \left( {\frac{\alpha ZE}{\sqrt {m^2c^4 - E^2} }
- \frac{1}{2}} \right) = 0 \\
 \end{array}
\end{equation}

\noindent
It gives two positive energy eigenvalues
\begin{equation}
\label{eq20}
\begin{array}{l}
 E = \frac{mc^2}{\sqrt {1 + \frac{\alpha ^2Z^2}{\left[ {\sqrt {\left( {j + 1
\mathord{\left/ {\vphantom {1 2}} \right. \kern-\nulldelimiterspace} 2}
\right)^2 - \left( {\alpha Z} \right)^2} + n_r } \right]^2}} } \\
 E = \frac{mc^2}{\sqrt {1 + \frac{\alpha ^2Z^2}{\left[ {\sqrt {\left( {j + 1
\mathord{\left/ {\vphantom {1 2}} \right. \kern-\nulldelimiterspace} 2}
\right)^2 - \left( {\alpha Z} \right)^2} + n_r + 1} \right]^2}} } \\
 \end{array}
\end{equation}

By all appearances, the two energy eigenvalues can not give the same set for
the same quantum system in Coulomb field. Their values of energy for the
ground state are not equivalent. It violates the uniqueness of solution.
However, the original paper wrote the equations (\ref{eq10}) in one of forms so as
to only choose the first formula. It made a gloss that the distinguished
energy eigenvalues were recovered. Such method should not be selected in any
scientific paper.

\section{Eigenfunctions set of equations (\ref{eq3}) violates boundary condition}

Now we can write down the complete formal solutions of the first-odder Dirac
equations (\ref{eq1}) with the recurrence relations for the coefficients of the
corresponding power series, in which the eigenvalues $E$ are determined by
the self-contradictory expression (3.20. Qua the formal eigenfunctions set
of the original Dirac-Coulomb equations with the transformation of function
(\ref{eq2}), it is
\begin{equation}
\label{eq21}
\begin{array}{l}
 f = \frac{1}{2}\sqrt {\frac{mc^2 - E}{mc^2 + E}} \rho ^{ - 1 + \sqrt
{\left( {j + 1 \mathord{\left/ {\vphantom {1 2}} \right.
\kern-\nulldelimiterspace} 2} \right)^2 - \left( {\alpha Z} \right)^2} }e^{
- \frac{\rho }{2}}\left( {\sum\limits_{n = 0}^{n_r } {d_n } \rho ^n -
\sum\limits_{n = 0}^{n_r } {b_n } \rho ^n} \right) \\
 g = \frac{1}{2}\rho ^{ - 1 + \sqrt {\left( {j + 1 \mathord{\left/
{\vphantom {1 2}} \right. \kern-\nulldelimiterspace} 2} \right)^2 - \left(
{\alpha Z} \right)^2} }\left( {\sum\limits_{n = 0}^{n_r } {d_n } \rho ^n +
\sum\limits_{n = 0}^{n_r } {b_n } \rho ^n} \right) \\
 \end{array}
\end{equation}

\noindent
When $j = 1 \mathord{\left/ {\vphantom {1 2}} \right.
\kern-\nulldelimiterspace} 2$, at the origin of the coordinate
system, it becomes
\begin{equation}
\label{eq22}
\begin{array}{l}
 \mathop {\lim }\limits_{\rho \to 0} f = \mathop {\lim }\limits_{\rho \to 0}
\frac{1}{2}\left( {d_0 - b_0 } \right)\sqrt {\frac{mc^2 - E}{mc^2 + E}} \rho
^{ - 1 + \sqrt {1 - \left( {\alpha Z} \right)^2} }e^{ - \frac{\rho }{2}} =
\infty \\
 \mathop {\lim }\limits_{\rho \to 0} g = \mathop {\lim }\limits_{\rho \to 0}
\frac{1}{2}\left( {d_0 + b_0 } \right)\rho ^{ - 1 + \sqrt {1 - \left(
{\alpha Z} \right)^2} } = \infty \\
 \end{array}
\end{equation}

\noindent The formal solution for the ground state is divergent at
the origin of coordinate system.

On the surface, this divergence can be explained that it has no mathematical
and physical consequences, because the equations (\ref{eq1}) have been called that
it was only given outside of the atomic nucleus, and there does not exist
the case $r \to 0$. However, if we accept this explain, then how to choose
the positive signs for the index $e_\chi $ and $e_\sigma $ in (\ref{eq17})? The
divergence (\ref{eq22}) implies that the probability density of the electron
appearing around the atomic nucleus would rapidly increase as it near the
atomic nucleus. If the formal solution (\ref{eq21}) is the real wave function in
Coulomb field, then all of the electrons would fall into the atomic nucleus.
The atom would collapse to become neutron-like. It is not true! In fact,
considering the atomic nucleus has a finite radius, the boundary condition
would take the other form, which was called the exact boundary
condition\cite{Chen:2007}. Unexpectedly, the Dirac-Coulomb equations with
the exact boundary condition have different energy eigenvalues from the
Dirac formula of the energy levels.

Consequently, the formal solutions (\ref{eq21}) of the Dirac-Coulomb equation can
but be understand that it violates the boundary conditions. Of course, this
serious contradiction was also ignored by the original paper. In
mathematics, it is well known that any formal solution which violates the
condition for the determining solution must be not the real solution of the
corresponding differential equation. Because of the divergence, we conclude
that the necessary mathematical deductions (\ref{eq21}) are not the real solution
of the Dirac-Coulomb equation.

\section{Conclusions}

In a word, because the formal eigenvalues set violates the uniqueness of
solution and the complete formal solutions for the ground state of the
corresponding differential equations in the Coulomb field of second-order
are divergent at origin, the given formula of the energy levels of the
Dirac-Coulomb equation exterior to the nucleus in the mentioned paper is a
pseudo energy gienvalues set, and the given wave functions, which were
expressed in terms of confluent hypergeometric functions, are also the
pseudo solutions of the Driac equation for the hydrogen and hydrogen-like
atom. Such kind of pseudo solutions of the Dirac equations indicate that in
relativistic quantum it is very important for some details being treated,
such as the boundary conditions, the uniqueness of solution, the existence
of eseigensolutions and the harmonic analysis and their homogenus spaces and
so on\cite{Heinrich:2006}.

For those incorrect deductions which violate the mathematical operation
rules, in their papers and the quantum mechanics textbooks, some authors
even called that what they related was only the physics but not mathematics,
and introduced some strange words such as ``mild divergence'' and
``decoupling procedure'' and so on, which are independent of any
mathematical and physical logic. However, for the quantum mechanics, once a
wave equation with the certain boundary condition or initial value condition
is introduced, all remnant physics should be essentially the pure
mathematics. At least, anyone cannot coin some mathematic formulas for those
important physics problems by using multitudinous wrong mathematics
calculations. Because those hidden pseudo solutions have brought the
comprehensive infection, some new correct deductions of the theoretical
physics are very difficult to be known. One should disclose such kinds of
pseudo theories in physics, although it is a work\cite{Robert:2002} for
lollygagging. For the Dirac equation in the Coulomb field, there are many
problems\cite{Seevinck:2001}\cite{Chen:2}, in particular the localization
problem and the Klein paradox\cite{Thaller:1992}, which need to be solved.

\end{document}